\documentclass[a4paper, 5p, twocolumn]{elsarticle}

\usepackage{lineno}
\usepackage{amssymb}
\usepackage{amsmath}
\usepackage{graphicx}% Include figure files
\usepackage{tabularx,array,booktabs}
\usepackage{url}
\usepackage{hyperref}
\usepackage[export]{adjustbox}

\usepackage[rightcaption]{sidecap}
\sidecaptionvpos{figure}{t}
\usepackage{todonotes}

\journal{Elsevier}

\begin{document}

\begin{frontmatter}

\title{Identifying drivers and mitigators for congestion and redispatch in the German electric power system with explainable AI}

\author[iek10,col]{Maurizio Titz\corref{cor}}
\ead{m.titz@fz-juelich.de}
\author[iek10,col,kit]{Sebastian P\"utz}
\author[iek10,col]{Dirk Witthaut}
\ead{d.witthaut@fz-juelich.de}

\address[iek10]{Forschungszentrum J\"ulich, Institute of Energy and Climate Research -- Energy Systems Engineering (IEK-10), 52428 J\"ulich, Germany}
\address[col]{University of Cologne, Institute for Theoretical Physics, Z\"ulpicher Str. 77,  50937 Cologne, Germany}
\address[kit]{Karlsruhe Institute of Technology, Institute for Automation and Applied Informatics, Hermann-von-Helmholtz-Platz 1, 76344 Eggenstein-Leopoldshafen, Germany
}
               
\cortext[cor]{Corresponding author}

\date{\today}

\begin{abstract}
The transition to a sustainable energy supply challenges the operation of electric power systems in manifold ways. Transmission grid loads increase as wind and solar power are often installed far away from the consumers. In extreme cases, system operators must intervene via countertrading or redispatch to ensure grid stability. 
In this article, we provide a data-driven analysis of congestion in the German transmission grid. We develop an explainable machine learning model to predict the volume of redispatch and countertrade on an hourly basis. The model reveals factors that drive or mitigate grid congestion and quantifies their impact. We show that, as expected, wind power generation is the main driver, but hydropower and cross-border electricity trading also play an essential role. Solar power, on the other hand, has no mitigating effect. Our results suggest that a change to the market design would alleviate congestion.
\end{abstract}


\end{frontmatter}

%\maketitle


%\linenumbers

\section{Introduction}

The mitigation of climate change requires a comprehensive transition of the energy system towards renewable energy sources \cite{rogelj2015energy}.
Wind and solar power generation have shown tremendous growth over the last decades \cite{ipcc2022energy,creutzig2017underestimated} and have enormous potential for further development \cite{jacobson2011providing}.
Recent data shows that wind and solar power are generally cheaper than fossil fuel alternatives in terms of the levelized costs of electricity \cite{irena2021renewable}. Hence, a transition to renewable sources is not only possible but also cost effective~\cite{victoria2020early}.

The integration of renewable power sources into the existing electricity system remains a challenge, though. Wind and solar power generation depend on the weather and are thus intrinsically variable and uncertain. Hence, future power systems must include storage and backup infrastructures or sector coupling to cover periods of low renewable power generation \cite{elsner2015flexibilitatskonzepte,victoria2019sector}.

Furthermore, the overall electricity yield is highly location dependent. Wind power resources are determined by large-scale wind patterns and local terrain \cite{jung2020integration,emeis2018wind}, while solar power resources are mainly determined by degree of latitude \cite{ipcc2022energy}. In Europe, the best locations for wind turbines are found around the North Sea and the British Isles \cite{creutzig2014catching}. Power transmission from these regions to the customers remains a challenge as transmission capacities are limited~\cite{pesch2014impacts}. In the long term, comprehensive grid extensions are needed~\cite{rodriguez2014transmission,ENTSOE_TYNDP}, while in the short term, grid operators have to apply congestion management.
In Germany, one of the pioneers of the transition to renewable energy, the total costs of all power grid stability measures reached around 2.3 billion Euro in 2021~\cite{BundesnetzagenturNetzengpassmanagement}.

What factors promote or mitigate power grid congestion? Wind power is considered the main cause, but it is certainly not the only important factor. In this article, we present an empirical study of congestion in the German power grid, aiming to identify drivers and risks beyond wind power. We base our study on a public database containing all interventions in Germany by transmission system operators (TSOs) to manage congestion in the German grid. We develop a machine learning (ML) model that predicts the volume of redispatch and countertrade per hour from large-scale power system features such as renewable generation, electricity prices, or cross-border trading. Explainable artificial intelligence (XAI) methods are used to quantify the importance and dependencies of all features, thus identifying the key driving and mitigating factors. Our study reveals the important roles of European electricity trading and other renewable sources such as hydropower.

The article is organized as follows. We first review essential aspects of the German power system, the electricity markets and congestion management measures in section~\ref{sec:background}. Methods and data sources are described in section~\ref{sec:methods}. Section~\ref{sec:results} summarizes all results including the prediction performance of the developed methods, the feature importances and, most importantly, the inferred dependencies. We summarize our findings and discuss implications for the energy transition in section~\ref{sec:discussion}.

\section{Congestion and Redispatch: An Overview}
\label{sec:background}

\subsection{Power Generation and Transmission in Germany}

Germany is one of the pioneers of the transition to renewable energy sources \cite{curry2019germany,hake2015german}, despite having only mediocre natural conditions. In 2021, the aggregated wind and solar power capacity amounted to 64~GW and 66~GW respectively \cite{foederalerneuerbar}. As a consequence, Germany is facing challenges that are characteristic for the energy transition. For instance, renewable power generation is strongly fluctuating, and so are electricity market prices~\cite{han2022complexity}. 

\begin{figure*}[tb]
    \centering
    \includegraphics[width=0.242\linewidth]{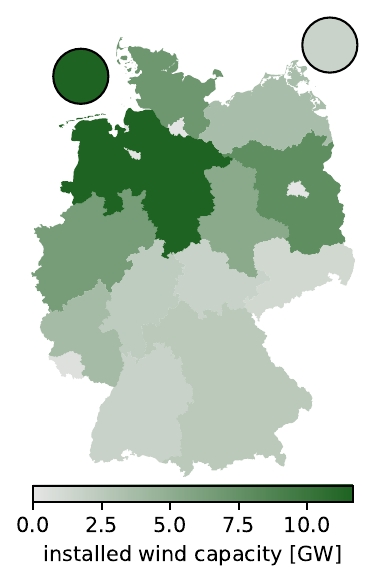}
    \includegraphics[width=0.242\linewidth]{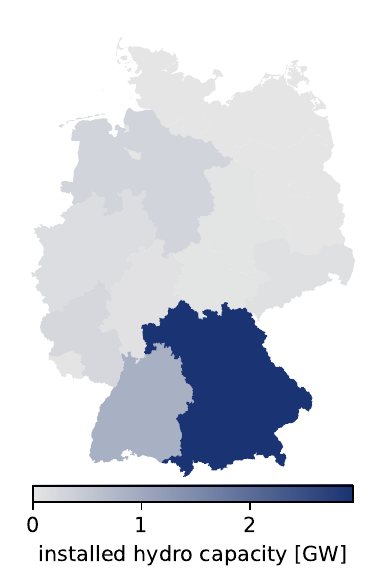}
    \includegraphics[width=0.242\linewidth]{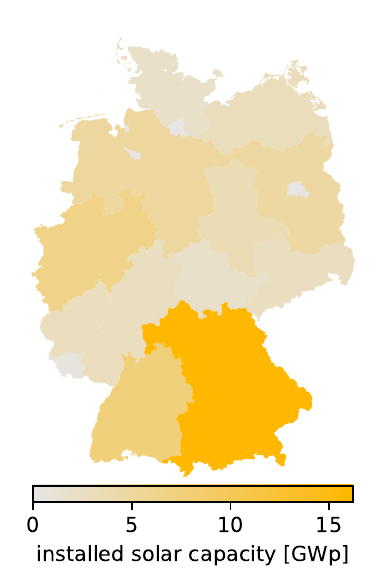}  \\      
    \includegraphics[width=0.242\linewidth]{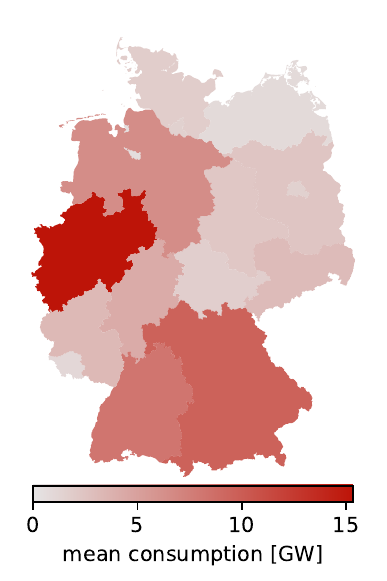}
    \includegraphics[width=0.242\linewidth]{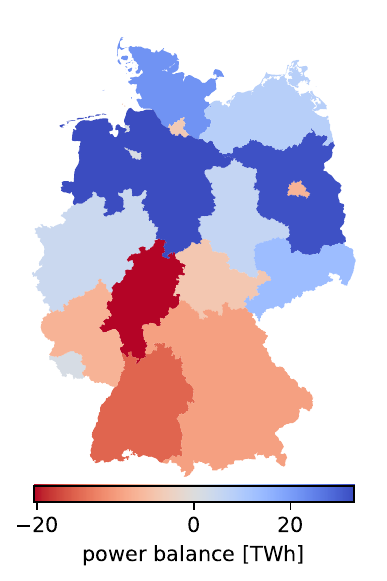}
    \includegraphics[width=0.242\linewidth]{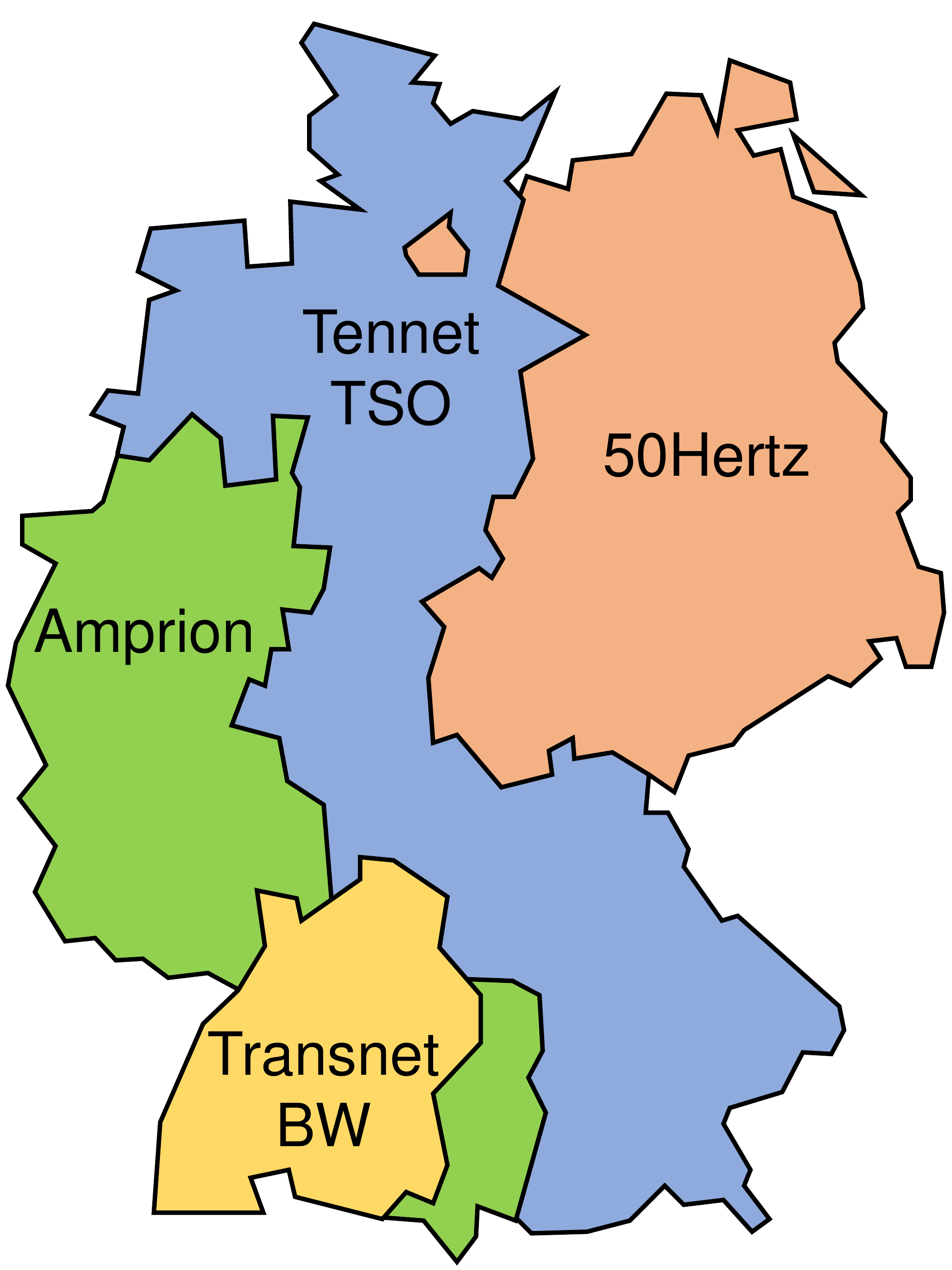}
    \caption{
    Overview of the German power system on the level of federal states.
    Top row: Installed generation capacity of wind power, run-of-river hydropower and
    solar PV. Circles represent accumulated capacity of offshore wind power in the North Sea and the Baltic Sea, respectively. 
    Bottom left: Mean power consumption.
    Bottom middle: Annual power balance, i.e. total generation minus consumption.
    Bottom right: The German power transmission grid is split into four control areas operated by different TSOs.
    Data has been obtained from~\cite{foederalerneuerbar}. Wind power data from 2022, hydro and PV data from 2021. Consumption and power balance data from 2019, except for Mecklenburg-Vorpommern (2018) and Saarland (2015). 
    }
    \label{fig:wind_cap_consumption}
\end{figure*}

A particular challenge arises from Germany's geographic properties. Favourable locations for wind turbines are located in the north and east of Germany~\cite{creutzig2014catching}, while several densely populated areas are located in the south and west. The uneven distribution of wind power capacity is further exacerbated by economic developments and political actions. In 2014, the federal state of Bavaria established a minimum distance rule that almost brought the development of wind energy to a standstill~\cite{langer2016qualitative}. At the same time, offshore wind power capacity has increased dramatically due to falling costs and the federal government has ambitious plans regarding further installations~\cite{bundesnetzagentur}.

The current situation in the German power system at the level of federal states is summarized in Fig.~\ref{fig:wind_cap_consumption}. Wind power capacity is concentrated in the federal states of Lower Saxony, Schleswig-Holstein and Brandenburg in the North, the East and offshore. In contrast, photovoltaic capacity is concentrated in the south, though the distribution is much more balanced than for wind generation. Hydropower is strongly concentrated in the southern federal states of Bavaria and Baden-Württemberg, but the overall capacity is much lower. The demand for electric power is highest in the densely populated North Rhine-Westfalia and in southern Germany, where industry is strong. As a result, wind power must be transported over long distances from eastern and northern Germany to southern and western Germany. 

However, power transmission capacities are limited and high levels of wind power generation often result in transmission grid congestion~\cite{bundesnetzagentur}. The extension of the transmission grid is therefore a central pillar of the decarbonization of the electric power system. Unfortunately, transmission grid extension is a complex challenge with high cost, and often lacks public acceptance~\cite{kuhne2018conflicts}. Due to their geographic distribution, PV and hydropower generation might lessen the transmission needs towards the South of Germany and thus alleviate congestion. 

For the reasons presented above, the German transmission grid is highly prone to congestion~\cite{pesch2014impacts}. 
Through investment essential properties of the power grid, such as the distribution of generation capacity or transmission capacity, can be changed to make congestions less likely. These changes are essential to keep the power system functional and efficient in the long run, but take time and will happen only given the right financial incentives.

On the operational time scale, too, congestions must be resolved at all cost.
Even if a congestion does not directly lead to thermal overloads, a congested power grid is overall more vulnerable to failures, in particular overload cascades. 
To limit the risk of malicious cascades, the \emph{N-1 rule} requires the power grid to be fully functional even if any one transmission line is lost \cite{wood2013power}. Transmission system operators (TSOs) are obligated to follow this rule and operate power generation such that the resulting power flows respect the N-1 rule. Different countries have adopted different strategies to solve this problem. In the following section, we describe the situation in the context of electricity grids and markets with a focus on Germany.

\subsection{Electricity Markets, Grids, and Congestion Management}

% European Electricity Markets and EUPHEMIA:
The synchronous European power grid spans a vast area from Portugal to Turkey. The dispatch of power plants is determined on electricity markets based on the offers and bids of the utility companies. To optimise the utilisation of available resources, a central algorithm called EUPHEMIA considers all bids in the whole European electricity market and calculates the best possible dispatch. EUPHEMIA is implemented to maximize ``the social welfare (consumer surplus + producer surplus + congestion rent across the regions) generated'' \cite{nemo2019euphemia,sleisz2014algorithmic}. The algorithm respects transmission capacity limits between bidding zones, but assumes unlimited transmission capacity within them, which is referred to as the ``copper plate model''. Bidding zones often, but not always, correspond to countries. For instance, Norway and Sweden consist of several bidding zones, whereas Germany and Luxembourg form a joint bidding zone.

% different categories of congestion management by time frame

\iffalse
% market does can not function properly
Frequent congestions indicate that the market mechanisms have no adequate internal representation of the relevant transmission limits. 
They might also 
In addition, the authors argue that the German imbalance pricing mechanism can give adverse price signals under network congestions. 
First of all, interventions induce direct costs in the form of compensation payments. Considering the whole grid, the (re)dispatch after an intervention will not be optimal, if the interventions just resolve congestions locally without taking into account the entire system. 
Generation may thus be more expensive and emissions may be higher. 
% cost distribution
Besides the increased total cost, the costs might also not be distributed fairly among the market participants. 
Typically, the extra costs have to be payed by the intervening TSO, which will pass them on to his customers. 
However, the TSO might not be responsible for the congestion. On top of that, he is not necessarily in the position to address the root cause of the problem in an efficient manner.
This means that the system fails to set market incentives for the problem to be solved efficiently.
\todo[inline]{The previous paragraph lacks any reference.}
\fi

\begin{figure}[tb]
    \includegraphics[width=0.49\linewidth]{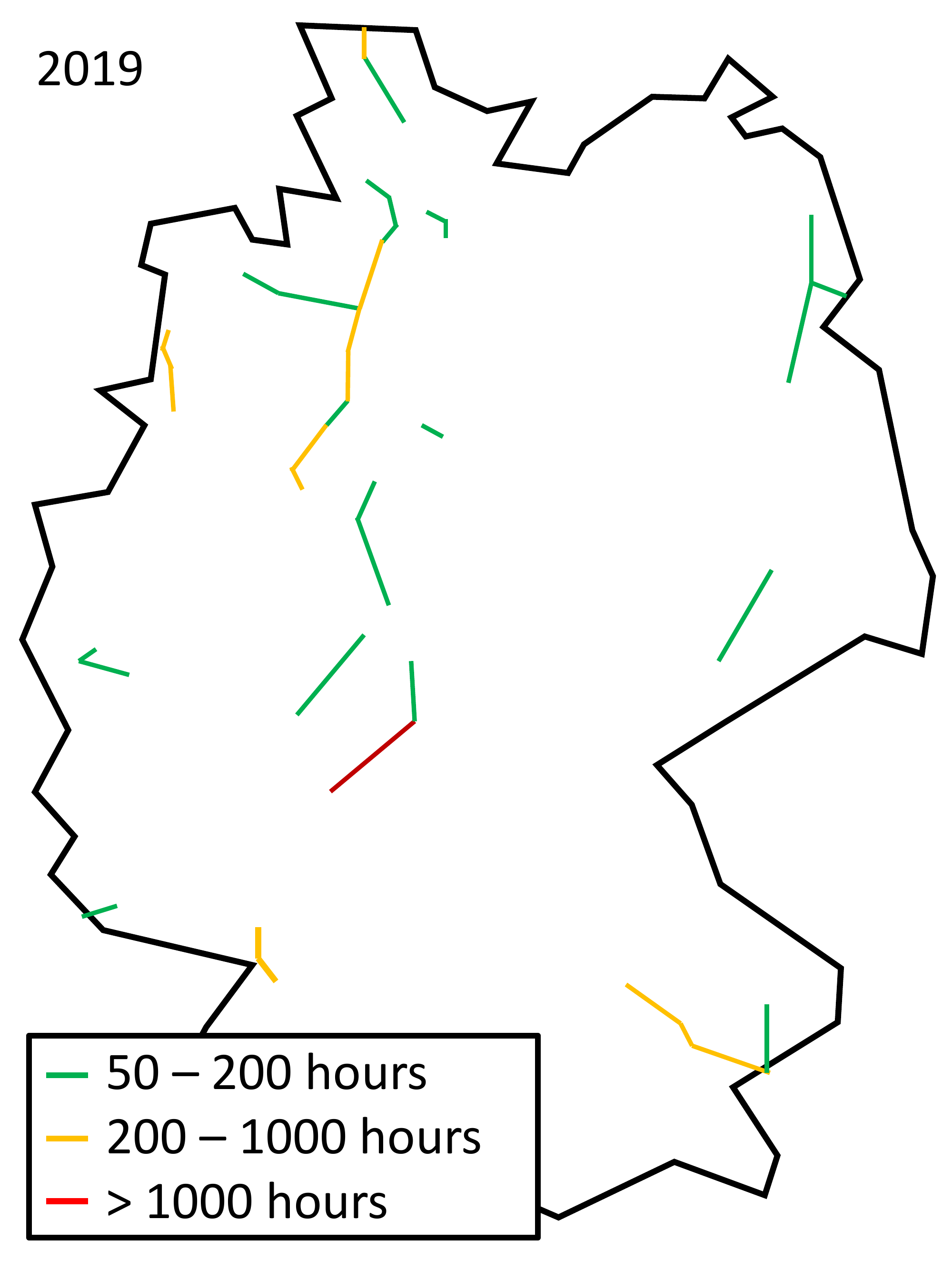}
    \includegraphics[width=0.49\linewidth]{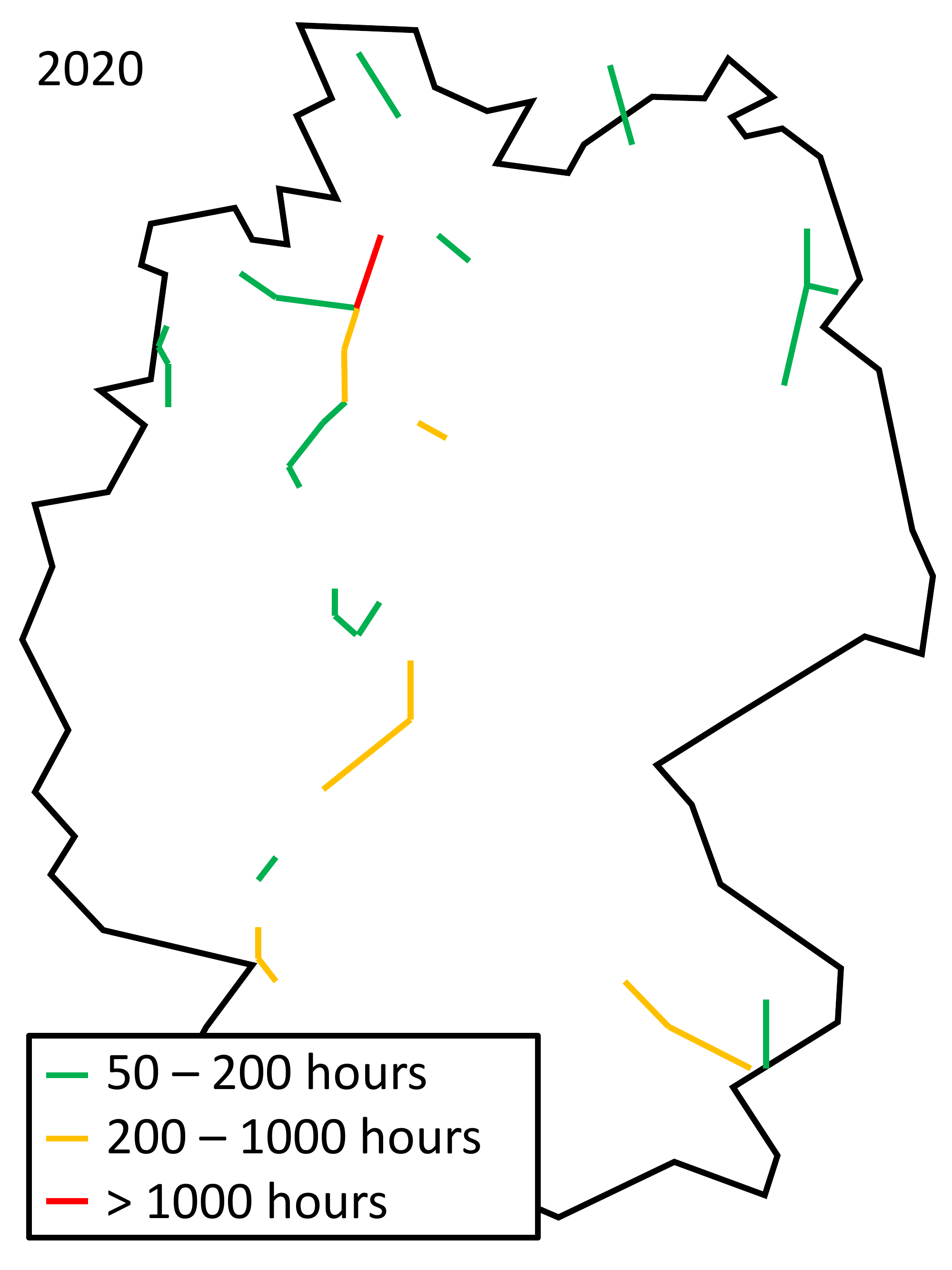}
    \caption{
    \label{fig:congestion_map}
    Congestion in the German power grid in 2019 and 2020. The maps show transmission lines that were congested in the sense that TSOs ordered a current-based redispatch. The color code shows the number of hours during which a redispatch measure was ordered.
    Figure based on \cite{bundesnetzagentur}.
    }    
\end{figure}

% Transmission limits: Bidding zone split vs. ex post congestion management
Some countries, such as Sweden, Norway, Denmark, have attempted to address the root cause of congestion by dividing the country into various bidding zones~\cite{spodniak2013area}. Hence, the limited transmission capacity between the bidding zones is explicitly represented in the EUPHEMIA algorithm. Nodal pricing has also been suggested as a solution, but it is currently not implemented in Europe~\cite{antonopoulos2020nodal}.
In contrast, all of Germany forms a bidding zone together with Luxembourg \cite{bundesnetzagentur}. Hence, transmission limits within Germany are not represented in the EUPHEMIA algorithm and German TSOs frequently have to perform congestion relief measures after the initial dispatch has been determined on the market. A map of congested lines is provided in Fig.~\ref{fig:congestion_map}.

% N-1 rule must be respected!
In spite of the problems laid out the German power system is highly reliable.
However, there are potential possible drawbacks regarding fairness and the setting of the right financial incentives associated with frequent congestion management \cite{chaves2014interplay}\cite{staudt2018predicting}. The need for congestion management could be decreased by dividing Germany into two or more bidding zones or even applying nodal pricing.

\begin{figure}[tb]
    \centering
    \includegraphics[clip,trim=0cm 27cm 0cm 0cm,width=1.0\linewidth]{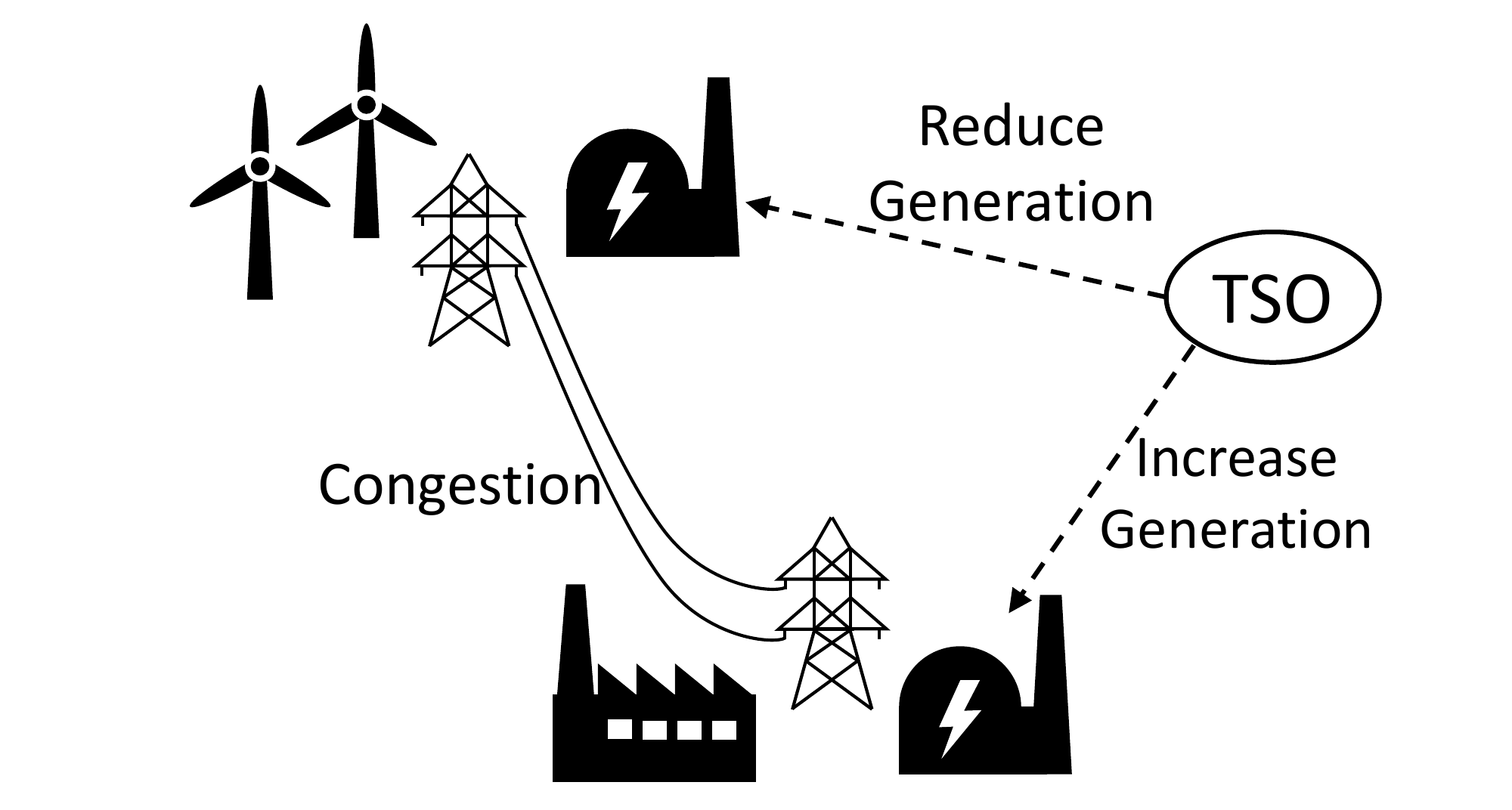}
    \caption{
    Schematic of congestion management.
    To reduce the load on a congested line the TSO orders a power plant lying upstream of the congestion to decrease its generation and a plant downstream of the congestion to increase generation. This way the congestion is alleviated without affecting power balance.}
    \label{fig:redispatch}
\end{figure}

\begin{figure*}[tb]
    \centering
    \includegraphics[width=0.5\linewidth]{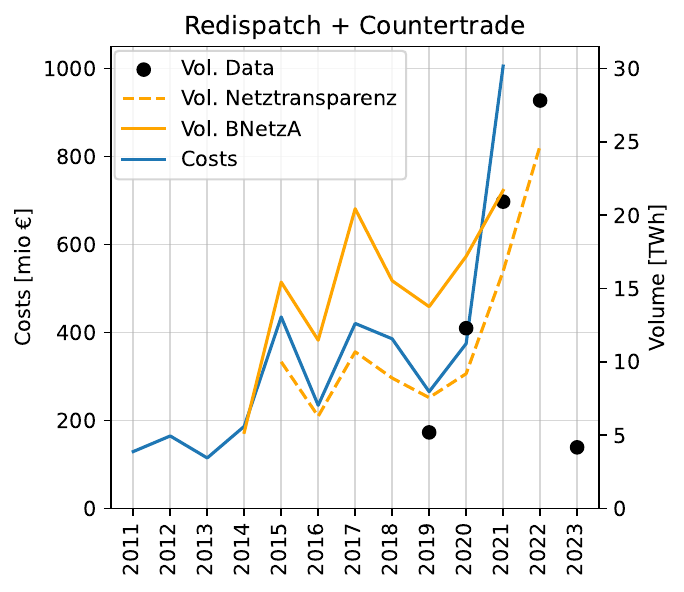}
    \vspace{0.04\linewidth}
    \includegraphics[width=0.45\linewidth]{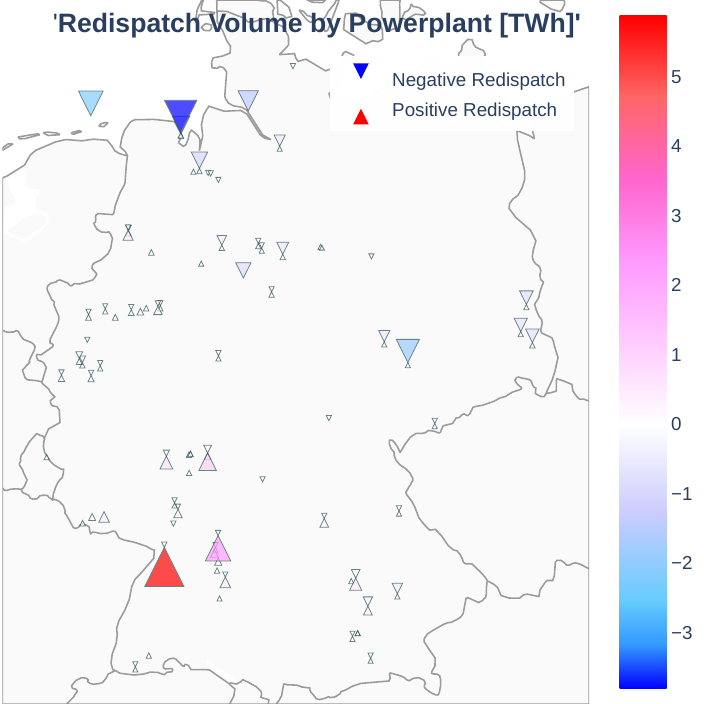}
    \caption{
    \label{fig:costs_congestion_map}
    Overview of redispatch in the German power system.
    Left: Development of redispatch and countertrade volume and costs\cite{bundesnetzagentur}. Volume data from \cite{bundesnetzagentur} which is complete, Netztransparenz\cite{netztransparenz_redispatch}, which lacks cross-border interventions, and data used for this work derived from the latter as explained in Sec.\ref{sec:target}. Note that only parts of the years 2019 and 2023 were analyzed, so that the volume in these years is very low for the latter. Right: Map of negative and positive redispatch volume per power plant in the time period considered in this work, data from \cite{netztransparenz_redispatch}. The negative redispatch takes place mostly in the north and east, positive in the south. Wind parks in the North Sea have been used for negative redispatch since the introduction of redispatch 2.0.
    }
\end{figure*}

% congestion relief measures: countertrading and redispatch
Preemptive congestion relief can be achieved by adjusting generation without altering the overall generation, thereby maintaining power balance, see Fig.~\ref{fig:redispatch}.
In countertrading, the TSO pays for generation de- and increases offered in a specific bidding zone on the intra-day market~\cite{meeus2020evolution}. The TSO has no influence or knowledge regarding which specific power plant will be affected~\cite{klos2020defining}. Accordingly countertrading is usually used when the congestion occurs close to the bidding zone border or is associated with cross-border flows. 
Redispatching on the other hand involves the TSO instructing a specific power plant operator to adjust generation and paying a predetermined compensation price \cite{bundesnetzagentur}.
By decreasing generation at one and increasing it at another specified power plant, congested transmission lines can be targeted more directly.
In Germany, redispatch is frequently observed as generation decrease in the northeast and increase in the southwest, as depicted in Fig.~\ref{fig:costs_congestion_map}. 

% einspeise management (curtailment): turn off renewables (wind)
Until recently for environmental reasons and because of their negligible marginal cost, renewable power plants were exempt from the aforementioned measures in Germany. Curtailment of renewable power generation (``Einspeisemanagement'' in German) is used as a last resort if a congestion cannot be resolved by other means~\cite{bundesnetzagentur}. As with redispatch, the operator of the renewable power plant is compensated for the energy that cannot be sold on the market due to curtailment. With the introduction of redispatch 2.0 and the abolition of the Einspeisemanagement, there is not separation between renewables and conventional power plants anymore, so that renewable generation can now be used in redispatch, too \cite{BDEW}. Notably, Germany did not establish a market for redispatch services although this has been recommended by the European Union \cite{martin2022strategic}.

% Add some details on redispatch in Germany
Congestion management has to be applied rather frequently in Germany. The total cost of all congestion management measures has increased to approximately 2.3 billion Euro in 2023, of which redispatch and countertrading have contributed approximately 1 billion Euro
\cite{bundesnetzagentur}, cf.~Fig.~\ref{fig:costs_congestion_map}.
Congested lines are found in several regions in Germany (Fig.~\ref{fig:congestion_map}), but some regions stand out: (i) areas in northern Germany with a high penetration of wind parks, (ii) North-south connections in the north and centre of Germany  and (iii) a connection to Austria.
The most frequently congested line in 2019 was the Dipperz - Großkrotzenburg line in Hassia (1052 hours). In 2020, the most frequently congested line was the Dollern - Sotrum line in Lower Saxony (1264 hours). In 2021, the high-voltage DC cable Kontek linking Denmark and East Germany was congested for 1959 hours, with Dollern - Sotrum ranking second at 1219 hours~\cite{bundesnetzagentur}.
A variety of power plants are used for redispatching, as shown in Fig.~\ref{fig:costs_congestion_map}, with the coal-fired power plant in Wilhelmshafen and the Rheinhafen power plant in Karlsruhe providing the highest contribution to negative and positive redispatch, respectively.

\subsection{Context and Scope of this Study}

% The others only do model based research
Transmission limits and grid congestion are major challenges for the integration of renewable energy sources. Scientific research on grid congestion is typically based on simulation and optimization models. For instance, large-scale energy system models typically include transmission limits as constraints (see, e.g.~\cite{horsch2018pypsa}). Advanced models jointly optimize the extension of generation and transmission infrastructures~\cite{neumann2022assessments}. A detailed model-based analysis of congestion in the German power system was presented by Pesch et al. combining a power plant dispatch model and a high-resolution transmission grid model~\cite{pesch2014impacts}.

% Empiric Research. No data on congestion. Hence study redispatch.
In this article, we adopt a complementary empiric approach.  As there is no publicly available data on congested lines, we cannot study congestion directly. We therefore analyze redispatch and countertrade data which have to be made available by law~\cite{netztransparenz_redispatch}, and together provide a good proxy for congestion.  While Einspeisemanagement is also an important tool for congestion management we did not included it in our analysis, as data is not readily available for all of Germany. At the same time we note that congestions that are managed by curtailing renewable power generation is by definition caused by an over supply thereof, and would thus not lead to new insights.

% short literature review
Empiric studies on congestion, redispatch and its causes are sparse. Staudt et al.~\cite{staudt2018predicting} employed various machine learning approaches to predict redispatch at the power plant level. However, beyond a cursory examination grounded solely in correlation, the authors did not analyze the underlying causes of redispatch. Wohland et al. \cite{wohland2018natural} have discussed the role of natural wind power variability on redispatch in Germany in the light of the public discussion. Monforti-Ferrario and Blanco analyzed impacts of congestion relief measures on air pollutants and greenhouse gas emissions \cite{monforti2021impact}. An empiric economic study at coarse scales can be found in \cite{kunz2016coordinating}.

% Aim of this paper
In this study, we apply eXplainable Artificial Intelligence (XAI) \cite{roscher2020explainable, machlev2022explainable} to identify the key factors contributing to congestion in the German transmission grid. Specifically, we aim to address the following questions: Besides wind, are there any other factors that contribute to congestion? Are there mitigating factors that reduce the negative impact of wind generation on congestion? 

% Why XAI?
Our XAI approach has several advantages over comparable methods of data analysis. Modern machine learning models can describe arbitrary nonlinear relations and interactions and thus go far beyond univariate studies or a linear correlation analysis. Feature attribution methods quantify the contribution of each feature \emph{without} being limited to the ceteris paribus assumption of classical sensitivity analysis. Furthermore, these methods provide a consistent measure of the importance of each input feature \cite{lundberg2020local}.

\section{Methods and Data}
\label{sec:methods}

\begin{figure*}[tb]
    \centering
\includegraphics[width=0.9\textwidth]{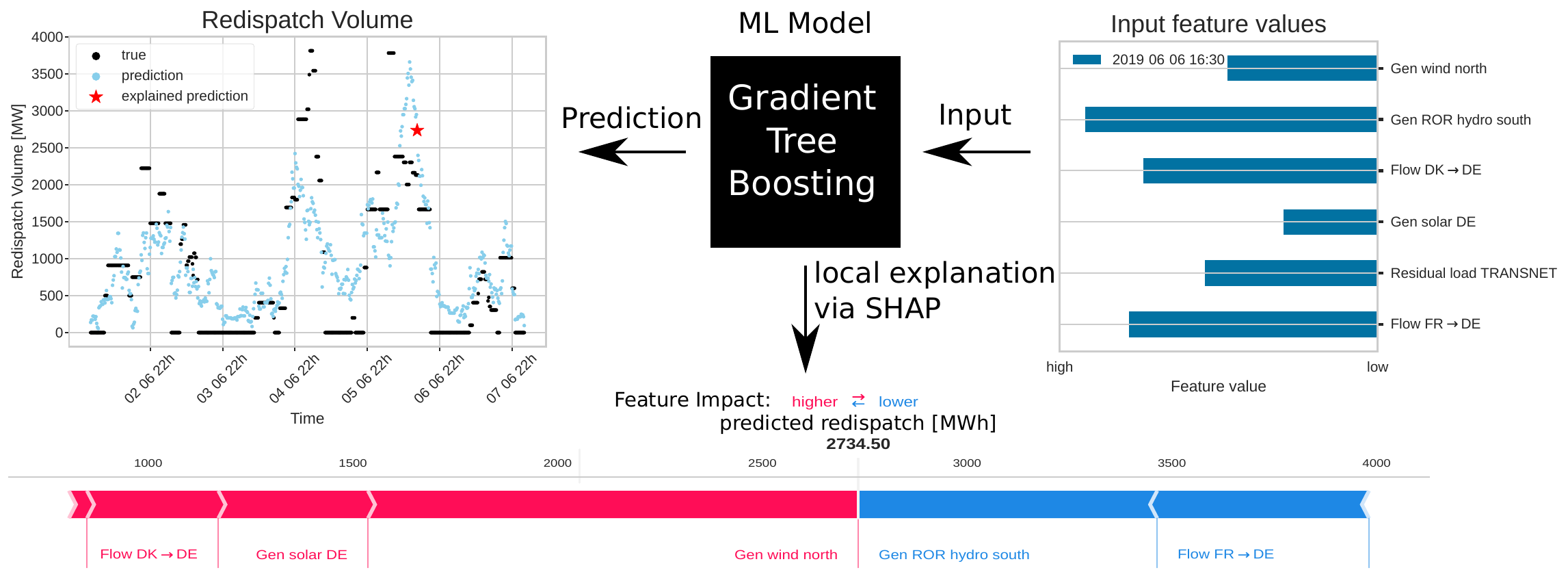}
    \caption{Schema of our approach. We train a machine learning model to predict redispatch volume in Germany from power grid features. We then use explainable artificial intelligence methods (SHAP) to gain insight into the relations the model learned. This way we identify drivers and mitigators of congestion in the German transmission grid.}
    \label{fig:my_label}
\end{figure*}

We develop an explainable machine learning model to predict the occurrence of redispatch in the German transmission grid. Our prime interest is the analysis of historical data to identify the main driving and mitigating factors for congestion and redispatch. The model may also be used in forecasting applications, but this is not the focus of the current work. A schematic of our approach is shown in Fig.~\ref{fig:my_label}.

\subsection{Redispatch Data}\label{sec:target}

% What data is available? % What do we do with it?
Data on redispatch and countertrade events in Germany is publicly available at \emph{netztransparenz.de}~\cite{netztransparenz_redispatch}. TSOs can also use power plants from the ``Netzreserve'' (grid reserve) for congestion management, which is included in the database. The database features all individual interventions including the start and end time, the requesting TSO and, for redispatch measures, the affected power plant.

% No geographic info available, thus focus on aggregated data 
The database does \emph{not} include information about the identity of the congested transmission line. Furthermore, a significant portion of the interventions is requested by several TSOs. Hence, it is mostly impossible to attribute a congestion event to a specific control area, let alone a specific line. We therefore model only the accumulated redispatch volume in Germany. 

% Amend data on cross-border redispatch
Entries on cross-border redispatch and countertrade are generally incomplete, as only interventions on the German side have to be publicised. To complete the target data, we thus assume an unreported intervention in equal size and opposing direction for every countertrade in the data. We note that this assumption is an approximation, only, as other schemes of intervention are possible in principle. For instance, a countertrade within Germany can be balanced by redispatch within Germany, or a redispatch in Germany can be balanced by a coutertrade in a neighboring country. Both possibilities cannot be inferred from the data alone, such that an approximation is necessary. The approximate completion scheme leads to an increase in the performance of the prediction model, although the difference is very small and the results are overall barely affected. 

% exclude: Congestion in other grids and voltage issues
As we focus on transmission limits and congestion in Germany, we exclude two types of interventions.
First, the dataset also includes interventions which were requested solely by foreign TSOs. As these were not used to relieve congestions in the German transmission grid, we discard them. Second, we discard interventions due to potential voltage limit violations and keep only current-related interventions.

% Target data: Aggregated positive AND negative volume
In principle, congestion management should not influence the power balance such that positive and negative redispatch and countertrade should be balanced. However, deviations regularly occur in practice because of several reasons. For instance, positive redispatch is also used to compensate for unexpected events such as unplanned power plant unavailability~\cite{staudt2018predicting}. To even out such occurrences and possible faulty or missing data, we aggregate negative and positive interventions. That is, the prediction target is given by the sum of the magnitudes of all interventions. This approach is further legitimised by a performance increase.

% Time Period considered:
We limit our analysis to the time period from May 2019 to January 2023 with a temporal resolution of one hour. Electrical power systems are ever evolving with transmission lines being constructed and generation facilities being commissioned or decommissioned. Major changes in the German power grid are the commissioning of the ``Thüringer Strombrücke'' in September 2017, the bidding zone split between Germany-Luxembourg and Austria in October 2018 and the decommissioning of three nuclear power plants in April 2023. All events lie shortly before or after the considered time frame. 
In 2019, the Netzausbaubeschleuningungsgesetz (``grid extension law'') was passed aiming to include smaller generating units in the redispatch process \cite{BDEW}. The \textit{redispatch 2.0} has been in full operation since June 2022 \cite{BundesnetzagenturNetzengpassmanagement} after a three months test phase. This change is visible by the presence of renewable-energy power plants in the data of the last months.

% Data quality
We finally note that the quality of the dataset is far from optimal. We find that the accumulated volume of all redispatch events from the dataset~\cite{netztransparenz_redispatch} deviates significantly from volume reported by the German regulating bodies~\cite{bundesnetzagentur}. This can probably be attributed to the non-availability of cross-border interventions in data from \cite{netztransparenz_redispatch}. Both numbers are compared in Fig.~\ref{fig:costs_congestion_map}, showing that the accumulated volume is smaller than the reported volume. After inferring missing cross-border interventions the difference is still significant for 2020, but negligible for 2021. Note that for years 2020 and 2023 we analyze only a part of the year, so that the shown total volume is much lower than in 2021 and 2022. Data on countertrading and redispatch per hour is also available from the ENTSO-E Transparency Platform\cite{ENTSOETransparencyPlatform}, but the discrepancy to the reported yearly volume is even higher.

\subsection{Features}
\label{sec:features}

% \usepackage{graphicx}
\begin{table}[tb]
\centering
\caption{Table of features used for redispatch prediction. All features are day-ahead features. Features in the upper row are available for each of the four German TSOs. Wind north and hydro south generation were derived by aggregating, as explained in section \ref{sec:features}. Features in the bottom row from all Germany's neighbouring countries were used, if available. Features marked by $^1$ were used in the reduced model, those marked by $^2$ in the full model.
}
\vspace{3ex}
\label{tab:my-table}
\resizebox{\linewidth}{!}{%
\begin{tabular}{l|l}
\multicolumn{1}{c|}{\textbf{Control Areas (DE)}} & \multicolumn{1}{c}{\textbf{Country (Neighbours)}} \\[0.7ex] \hline\hline
Gen Wind (on$^2$, off$^2$, total$^1$) & \multicolumn{1}{l}{Cross-Border Flows with DE$^{12}$} \\ 
Gen Solar$^1$ & \multicolumn{1}{l}{Price$^{12}$} \\ 
Gen ROR Hydro$^2$ & \multicolumn{1}{l}{Price Diff DE$^{12}$} \\ 
Gen Rest$^{12}$ &  \\ 
Load$^2$ &  \\ 
Residual Load$^1$ &  \\ 
Gen Wind North$^1$ &  \\ 
Gen ROR Hydro South$^1$ &  \\ 
\end{tabular}%
}
\end{table}

% We use only day-ahead forecasts as features.
We use a variety of features from power system and electricity market operation as inputs for our machine learning models. All features used in our model are day-ahead forecasts, because the EUPHEMIA algorithm calculates the dispatch on a day-ahead basis. As a consequence, redispatch measures are also planned primarily on the basis of day-ahead forecasts. 
Actual values of generation and cross-border flows are not included in the model. These values already include changes due to congestion management impeding any causal interpretation.

% Base features: download from ENTSO-E, complete list.
All input data is gathered from the ENTSO-E Transparency Platform~\cite{ENTSOETransparencyPlatform}. We use three classes of base features: (i) the load, wind generation, solar generation, run off river (ROR) hydro generation and the remaining (dispatchable) generation for each control area plus the offshore wind generation in the North Sea and the Baltic Sea, (ii) the scheduled cross-border flows between Germany and all its neighbouring countries and (iii) the electricity prices in Germany-Luxembourg and in all its neighbouring bidding zones as well as the respective price differences with regards to Germany-Luxembourg.

% engineered features: residual load, wind north hydro south
In addition, we engineer further features to improve the interpretability of the developed models. The residual load in a control area is obtained by substracting the non-dispatchable renewable generation from the load. Furthermore, we define two proxies for the total wind power generation in the North of Germany and the total ROR generation in the South of Germany. Unfortunately, data on the level of control areas does not lend itself to extracting features for the North and South directly, since the Tennet area spans the whole length of Germany (Fig.~\ref{fig:wind_cap_consumption}). However, the geographical distribution of wind generation capacity is such that almost all wind generation in the Tennet area can be is located in the North of Germany. We thus define the aggregate wind generation in the North as the sum of the wind generation in the Tennet and 50Hertz control areas. The opposite is true for ROR hydro generation, which is located primarily in the South of Germany. We thus define the sum of hydro generation in the Tennet and Transnet control areas as a proxy for the hydro generation in the South. 

% Two models. Use either aggregated data or base data. 
We have developed and evaluated two ML models, one using only the base features and one optimized for explainability using also engineered features. When using the aggregated features ``Wind generation North'' and ``Hydropower South'', we exclude the corresponding base features to reduce redundancy and thus increase explainability. 
% market features

% exclude hydro 50Hertz:
As mentioned before, there is very little hydropower generation capacity in northern Germany. In the exploratory phase, hydropower generation in the 50Hertz control area however showed an unexpectedly and unreasonably high feature importance. We concluded that the model was fitting merely a correlation, not a causal relation, and therefore excluded the feature from all models. 

% Testing further features:  
Finally, we tested several other features with a potential impact on power system operation. Ambient conditions can affect the transmission capacity of lines equipped with dynamic line rating and thus the necessity for redispatch~\cite{blumberg2019impact}. We thus tested air temperature as a feature, but found no relevance and thus discarded the feature. Similarly, time and seasonality features showed no relevance and were not used in the models discussed later.

% Features are strongly correlated!
It is important to note that the features used are all correlated in one way or another and cannot be disentangled. Wind (Solar) generation in different control areas is obviously positively correlated. Solar and wind generation are negatively correlated due to opposite seasonal profiles, while solar generation and load are positively correlated due to a similar daily profile. Furthermore, almost all features are correlated through their interaction with the market. These correlations prevent a data analysis with direct univariate techniques or linear models. In contrast, advanced ML models can break down the effect of correlated features and reveal otherwise undetectable nonlinear effects and feature interactions using the methods described below.

\subsection{Machine Learning Methods}

% We use GBT and perform HPO via 5-fold CV using a random split.
We use gradient-boosted trees (GBT), which offer state-of-the-art performance at low computational costs while enabling a fast and efficient model explanation~\cite{lundberg2017unified}.
We perform hyperparameter optimization (HPO) using random search with 5-fold cross validation. We don't use a time series split since we are interested in explanation, not in forecasting. Instead, we use a group shuffle split with 24 hour gaps so that training and test data don't contain samples that include the same redispatches.

% Recursive Feature Elimination
Additionally, we perform Recursive Feature Elimination (RFE) to reduce model complexity and thus improve interpretability. In RFE one iteratively fits a model, determines the importance of the different features to the model output and eliminates the least important feature. Feature importances are determined via SHAP values as described below. We then select a model that shows performance close to the initial model while being much less complex. 

% We produce two models.
We fit models with different input features. The first model uses only the base features as defined in Sec.~\ref{sec:features}. This model is meant to take in all relevant data with minimal feature redundancy. The second model uses the engineered features while dropping the related features as explained above to improve explainability.

\subsection{SHAP values}
\label{sec:shap}

% Fundamentals of SHAP
GBTs are not inherently transparent but enable an efficient ex-post explanation via SHapley Additive exPlanations (SHAP)~\cite{lundberg2017unified}.
SHAP values quantify the impact of each input feature on the model prediction in a mathematically consistent way. In our case, SHAP values quantify how much a feature contributes to the predicted redispatch volume for the given input feature values. More precisely, if the model predict the redispatch volume $f$ from the feature values $x_1,\dots,x_n$ we have
\begin{equation}
    f(x_1,\dots,x_n) = \phi_0(f) + \sum_{j=1}^n \phi_j(f;x_1,\dots,x_n),
\end{equation}
where $\phi_j(f;x_1,\dots,x_n)$ denotes the SHAP values for the $j$th feature. These \emph{local} explanations make individual predictions interpretable. 
SHAP values are unique in fulfilling certain desirable properties for model explanation \cite{lundberg2017unified} and thus avoid inconsistencies present in other methods.

% From local to global model explanation
From the local explanation of individual predictions, one can derive a global understanding of the model via feature importance, dependence plots and interactions plots \cite{lundberg2020local}. The importance of the $j$th feature is obtained by aggregating its SHAP values over all samples $s$
\begin{equation}
    FI_j = \frac{1}{\mathcal{N}} \sum_{s} |\phi_j(f;x_1^{(s)},\dots,x_n^{(s)})|.
\end{equation}
The normalization factor $\mathcal{N}$ is chosen such that $\max_j FI_j = 1$. This global importance measure is used in the RFE. 

SHAP dependence plots (cf. top panel ~Fig.~\ref{fig:dependencies}) give detailed insight into how different feature values affect the model's output. In such a plot, the SHAP value $\phi_j$ is plotted versus the feature value $x_j$ for all samples $s$. Lastly, SHAP interaction values (cf. bottom panel ~Fig.~\ref{fig:dependencies}) quantify the impact of the interaction of two features~\cite{lundberg2020local}.

\section{Results}
\label{sec:results}

\subsection{Overview and Performance}

\begin{figure}[tb]
    \centering
    \includegraphics[width=\linewidth]{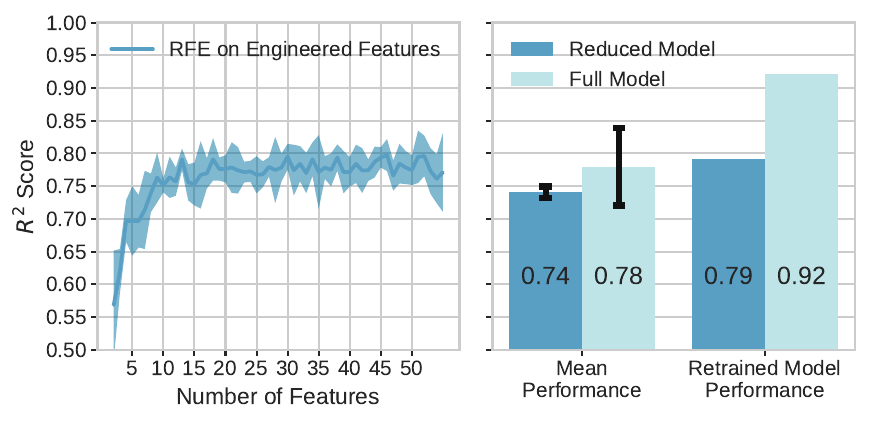}
    \caption{
    Performance of the developed ML models for the prediction of redispatch volume.
    Left: Performance during Recursive Feature Elimination (RFE) for the model including engineered features. By performing further manual feature selection we obtained a reduced model that performs well and is explainable. Right: Performance of the reduced model and the full model which does not contain engineered features. The mean cross validation performance is significantly lower for both models because the retraining was performed on a larger training set. 
    %The model was retrained on the whole data set used for the cv before and tested on a disjoint test set. 
    The reduced models mean performance is slightly lower. This is mostly due to the fact that we eliminated features that captured only correlational relations but no causality.}
    \label{fig:rfe_score}
\end{figure}

% Overview: Two models
We will now present two models that were trained on different input features. The \emph{full model} was trained on all base features but no engineered features. For the \emph{reduced model} we started with a feature set that includes the engineered features but discards the correlated ones, as explained before. We then performed recursive feature elimination (cf.~Fig.~\ref{fig:rfe_score}) and manual feature selection to obtain a model that performs well and can be explained more easily.

% Performance
Both models show a decent performance considering the course grained nature of the input features and the quality of the target data (Fig.~\ref{fig:rfe_score}). The $R^2$ score reaches $0.74$ and $0.78$ when averaged over the cross validation sets and $0.79$ and $0.92$ for the retrained models. The retrained performance is higher because the respective models are trained on a larger training set. The high score of the full model is partly due to high variance of the full models and should not be overrated.

% RFE and Manual Feature Selection Details
The recursive feature elimination procedure shows that a decent model performance can be obtained already for a rather small number of features. We choose six features and tune the feature set manually. In particular, we find very similar model performance if we include either the physical cross-border flows or the price difference between two countries. In these cases, we choose the cross-border flows for better interpretability. 

% Finally: The nasty bit about the Czech electricity prices.
Finally, RFE always leads to a model including the Czech electricity prices. A closer inspection of the data shows that the Czech price has consistently been higher during the last months of the interval of interest. Replacing it with a rolling average did not impact performance significantly. We thus conclude that the model used the Czech electricity prices primarily to identify a certain period in time with overall high redispatch volume. Hence, we assume that this feature does not contain any relevant causal relation and eliminated it from the model. Most of the performance difference to the full model stems from doing so.
%The performance loss from doing so is also responsible for the majority of 

\subsection{The reduced Model}

\subsubsection{Overview and Feature Importances}

\begin{SCfigure}
    \includegraphics[height=4.2cm]{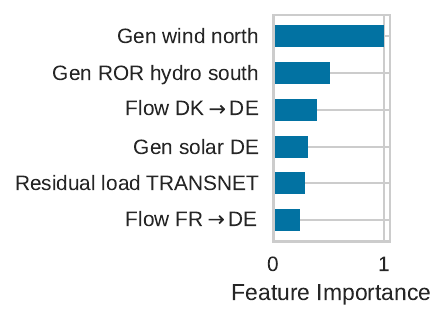}
  \caption{\protect\rule{0ex}{5ex}Feature Importance in the reduced model. As expected wind generation in northern Germany is the most important factor.}
  \label{fig:feature_importance}
\end{SCfigure}

\begin{figure*}[tb]
    \centering
    \includegraphics[width=0.99\textwidth]{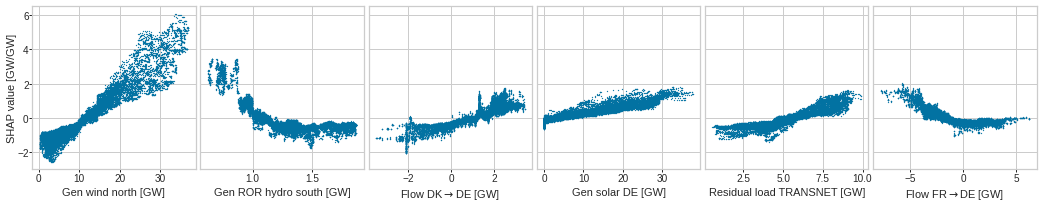}
    \includegraphics[width=0.99\textwidth]{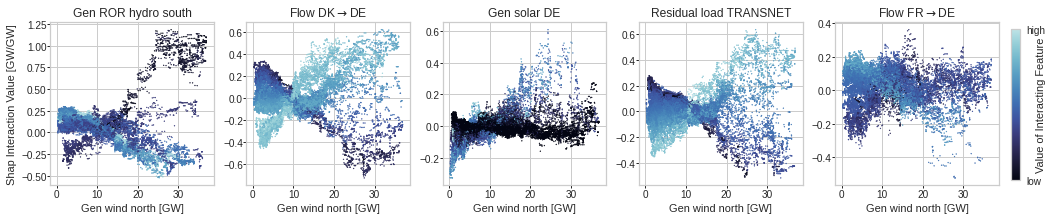}
    \caption{
    Dependencies and Interaction in the reduced model.
    Top: SHAP dependence plots for all features in the reduced model. As expected higher wind generation leads to more congestion. So do imports from Denmark and high residual loads in the Transnet control area. Surprisingly, high solar generation is also related to increased congestion. Hydro generation in the south and imports from France decrease congestions.
    Bottom: SHAP interactions plots for the most important feature, the wind power generation in northern Germany. 
    All other features besides solar generation show a systematic interaction with the wind generation. Hydro generation, exports to Denmark, low residual loads and Imports from France all mitigate congestion caused by high wind generation.}
    \label{fig:dependencies}
\end{figure*}

The final reduced model relies on six features, whose overall feature importance is summarized in  Fig.~\ref{fig:feature_importance}.
As expected, wind power generation in northern Germany is the most important feature. Remarkably, run-of-river hydropower generation in southern Germany ranks second. The cross-border flows to Denmark and France rank third and sixth, respectively, showing that the international electricity market is an important factor for congestion in the German transmission grid. Solar power generation and the residual load in the Transnet control area rank fourth and fifth, respectively, Notably, five out of six features relate to either the North or South of Germany, i.e. one side of the major transmission grid bottleneck. 
We will now discuss the role of these features in detail on the basis of SHAP dependence and interaction plots shown in  Fig.~\ref{fig:dependencies}.

\subsubsection{Wind Power Generation}

Wind power is the most important driving factor for congestion in the German transmission grid. The SHAP analysis confirms the expectation formulated in section~\ref{sec:background}. Wind power generation in northern Germany is the most important feature. The dependence is approximately linear with strong dispersion for low and high generation values. This dispersion can be partly explained by feature interactions as discussed below.

\subsubsection{Cross-Border Flows}

% The opposite role of of Denmark and France.
The cross-border flows with France and Denmark impact the model output in opposite ways: Imports from Denmark increase congestion, while imports from France alleviate them. This can be explained by the different positions of Denmark and France with respect to the Northeast-Southwest bottleneck in the German transmission grid.
Obviously, imports from Denmark will lead to congestion whenever they are to be consumed in southern Germany or exported to other countries in southern Europe. 
The connections to France lie on the other side of the bottleneck, such that imports from France can cover demands in the southwest without passing through it. Exports to France covered by generation in the north or east of Germany must go through the bottleneck, increasing the likelihood of congestion. 

% Interactions 
The interaction plot suggests, that exports to Denmark and imports from France actually alleviate the negative impact of high wind generation. While the former appears to be a causal connection, as the power imported from Denmark directly causes congestion, the latter is probably just correlational. Imports from France don't alleviate the congestion directly, but when power is imported from France, less power has to be transmitted within Germany.

\begin{figure*}[tb]
    \centering
    \includegraphics[height=4.5cm]{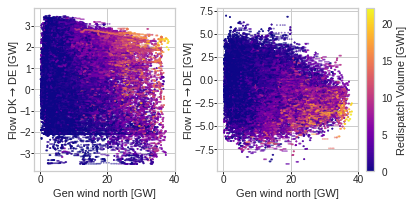}
    \includegraphics[height=4.5cm]{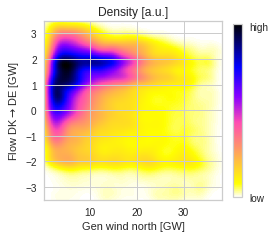}
    \caption{
    Raw data analysis of the relation of redispatch and cross-border flows.
    Left: Redispatch volume as a function of wind generation and cross-border flow from Denmark (France). Clearly high imports from Denmark on their own do not necessarily lead to congestions. However, when there is significant wind generation in Germany, these imports substantially exacerbate the problem.
    The same goes for exports to France.
    Right: Kernel density estimation for imports from Denmark and wind generation in northern Germany. Even for high wind generation Germany often imports electricity from Denmark.
    }
    \label{fig:wind_flows_vol}
\end{figure*}

% Discussion of Denmark in the Context of Euphemia: Denmark may intensify problems.
The importance of cross-border flows suggests that the international dispatch has significant impact on congestion within Germany. The EUPHEMIA algorithm takes into account capacity limits between bidding zones but ignores transmission capacity limits within Germany. Hence, it can schedule cross-border flows that intensify congestion within Germany. If for instance, Denmark generates a lot of wind power, it might be transported to southern Europe via Germany, even though German north-south transmission capacity is already utilized by German wind generation. This is also evident when plotting the redispatch volume as a function of wind generation and Denmark cross-border flow, see Fig.~\ref{fig:wind_flows_vol}. Once again, the opposite can be observed for France in the equivalent plot.

% No clear correlation of wind power generation and electricity imports from DK.
From a congestion management perspective, Germany would not import any electricity from Denmark during times of high wind generation. From an economic viewpoint, one might expect the opposite. Denmark has an even higher share of wind power than Germany, with a generation capacity exceeding the average grid load. Furthermore, wind power generation in Denmark and Germany are strongly correlated~\cite{olauson2016correlation}. Hence, there are strong incentives to export power from Denmark to Germany in times of high generation~\cite{nycander2020curtailment}.
In fact, the data analysis shows no negative correlation between wind power generation and imports from Denmark. In hours with low wind power generation, imports to Germany are much more likely than exports. In hours with wind generation exceeding 30~GW, we regularly find both cases with imports and exports. That is, Germany does clearly not stop importing electricity when producing large amounts of wind power.

Cross-border flows to Austria and Switzerland would appear to be relevant to the problematic North-South flows but were not selected during recursive feature elimination. This is to be expected for Switzerland because flows are usually small and thus of little importance. It is surprising for Austria, though, because the average of the absolute cross-border flows is higher for Austria than for any other neighbours. Furthermore, there are several lines close to the Austrian border that are often congested (Fig.~\ref{fig:congestion_map}). 
We find two possible explanations. First, congestion events close to the Austrian border may be underrepresented in the target data as they are resolved through cross-border redispatch and countertrade. 
Second, the impact of the Austrian power sector on congestion in Germany may be partly captured through the remaining variables serving as proxies. We will discuss this aspect in the next section.

\subsubsection{Hydropower Generation}

% Overview and dependence.
Hydropower generation is the second most important feature and counteracts congestion. The SHAP dependence plot (Fig.~\ref{fig:dependencies}) shows a strong decrease of the redispatch volume up to a generation of approximately 1.2 GW and a saturation afterward.
The decreasing relation is to be expected as run-of-river hydropower is mostly installed in southern Germany (Fig.~\ref{fig:wind_cap_consumption}). A high hydropower generation thus reduces the demand for transmission from Northern Germany.

% surprising importance of hydro and its explanation
The magnitude of the dependence is surprising, though, considering the limited total capacity.
We note that hydropower generation in Southern Germany is correlated to hydropower generation in the alpine region in general. Especially Austria and Switzerland cover large parts of their total electricity demand from hydropower. Given the importance of international electricity trading, it appears reasonable that hydropower generation in these countries will have a significant impact on congestion in Germany. The hydro generation feature in the model may thus serve as a proxy for the overall hydropower generation in the alpine region. However, a comprehensive understanding remains difficult.

% Exclude seasonality feature
We further tested a possible coincidence effect as a possible reason for the high feature importance. Hydro generation has a clear yearly profile such that the dependence may be a coincidence encoding a seasonal profile of the target feature. To test this possibility, we replaced the hydro generation feature by a rolling average or a synthetic seasonality feature. In both cases, the performance dropped significantly. Hence, we refute the possibility of a mere seasonal coincidence and conclude that hydro generation does in fact have a real impact. 

\subsubsection{Solar Power Generation}

% General result is surprising!
The SHAP dependence plot shows that the redispatch volume generally increases with the solar power generation in Germany. This finding is surprising as solar photovoltaics is primarily installed in the south of Germany as shown in Fig.~\ref{fig:wind_cap_consumption}. One might thus expect that solar generation would reduce the need for transmission from North to South, but this is not the case.

\begin{figure*}[tbh]
    \centering
    \includegraphics[width=0.7\textwidth]{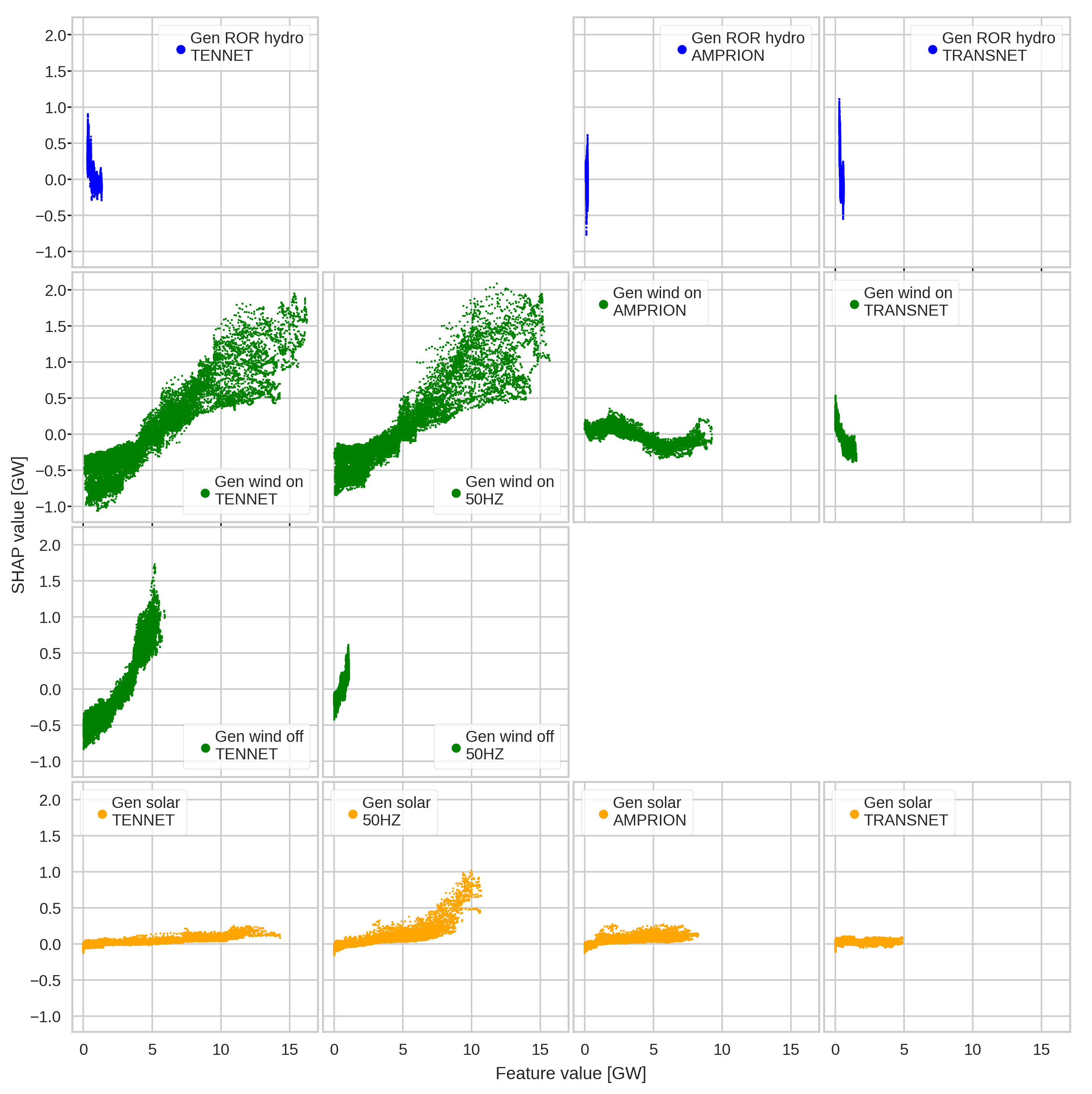}
    \caption{Dependence plots for renewable generation in all control areas in the full model. All dependence plots are plotted on the same scale to facilitate comparison.
    Hydro generation in the Tennet and Transnet areas shows an extremely strong negative correlations, whose magnitude appears to large to capture a causal relation. Wind generation in the North and offshore increases congestion, where offshore wind has a slightly stronger impact. Wind generation in the Transnet area is related to a decrease in congestion, while the relation is unclear for the Amprion area. 
    Solar generation overall has very little impact per generated power. Solar generation in the 50Hertz area is related to increasing congestion, while no systematic dependence is observed for the other control areas.
    }
    \label{fig:comparison_renewables}
\end{figure*}

% Test for different control areas?
To investigate potential regional effects, we consider alternative models, replacing the aggregated solar generation in Germany by the solar generation in the four control areas, respectively. Solar generation in the different control areas is strongly correlated however and we find a very similar performance in all cases, such that no further conclusions can be drawn at this point. We come back to this issue when discussing the full model in section \ref{sec:results-full-model}

% Test for daily 
We further tested whether the observed dependence is just a coincidence, similar to the case of hydropower. Solar generation has an obvious daily profile, such that the model may use this feature as a proxy for time of day. Replacing the solar generation by it daily profile, we find a significant decrease in the prediction performance. Hence, we refute the possibility of a mere daytime effect and conclude that solar power generation does in fact have a real impact.

\subsection{The full model}
\label{sec:results-full-model}

% Intro and performance.
We now turn to the second model for the redispatch volume -- the full model containing all base but no engineered features. As shown in Fig.~\ref{fig:rfe_score}, the mean performance of the model is slightly better than that of the reduced model. However, the model uses 42 features, such that the interpretation is more cumbersome.

% General results
We find that the feature importances and dependencies are generally consistent with the results obtained from the reduced model. In the following analysis we focus on the role of renewable power generation and the different control areas. In particular, we reconsider the open questions regarding the impact of solar and hydro generation raised in the preceding section. To enable a quantitative comparison we provide all SHAP dependence plots with the same axis scaling in Fig.~\ref{fig:comparison_renewables}. 

% Wind power
The relations of redispatch and wind power generation depend strongly on the location. Onshore wind in the Tennet and 50Hertz control area show similar dependencies: Redispatch increases almost linearly with the generation, showing a strong dispersion. The dependence is even stronger for offshore wind power generation in the two control areas. This comes as no surprise as \emph{all} offshore wind power has to pass through the same bottlenecks, while the average distance to the customers is even larger. 
Wind power in the Transnet area shows a reversed dependence, as the control area is located south of the bottlenecks. High generation reduces the need for transmission from the North and thus alleviates grid congestion. Wind power in the Amprion control area has a less clear dependence for geographic reasons.

% Solar Power
Solar power generation does not have a strong dependence, except for the 50Hertz region in northeastern Germany. In this region, we find a clear non-linear increase in redispatch volume with generation, as expected. In the Tennet and Amprion control areas we find a very weak positive relation. 
Remarkably, we do not find any correlation for the Transnet control area. This is surprising as we expect that generation in this region should relieve the grid similar to the case of wind power.

% hydropower
Hydropower shows a remarkably strong negative relation to the redispatch volume, especially for the Transnet control area. 
The disproportionate impact of hydro becomes particularly clear in Fig.~\ref{fig:comparison_renewables} in comparison to wind and solar power. Given the limited generation capacity, this dependence appears much too strong to be causal. This further supports our hypothesis that the learned relation between redispatch and hydropower generation stems from the great importance of hydropower in the Alpine region in general.

\section{Discussion and Conclusion}
\label{sec:discussion}

The decarbonization of the electricity system poses new challenges to the power grid. 
Higher grid loads make the power system more vulnerable and have to be addressed via costly congestion management. 
In Germany, congestions are mostly found along a north-south bottleneck in the transmission grid and are a result of the transmission need from the north which boasts high wind power generation to the south with its strongly negative power balance.

In this work, we analyzed the drivers and mitigators for congestion in the German transmission grid using eXplainable Artificial Intelligence (XAI). 
%Most congestions stem from the high transmission need from north to south Germany. 
To this end, we trained a gradient-boosted trees model to predict aggregated redispatch and countertrade volumes from day-ahead power grid features such as generation, load, price and cross-border flow data. By combining feature engineering, recursive feature reduction and manual feature selection we reduced the model complexity to six input features while achieving a mean $R^2$ score of $0.74$. In a second model, we removed all engineered features to reduce redundancy and achieved a comparable performance of $0.78$, now using 42 features. 
We then used SHapley Additive exPlanation (SHAP) values to interpret the models.

%The impact of all features can be described by the north-south transmission bottlenecks.
As expected, wind generation in northern Germany is the most important driving factor. 
Higher residual loads in the south, in our model represented by those in the Transnet control area, also increase the likelihood of congestions. 
Beyond these known driving factors, the model also allows us to identify \emph{mitigating} factors. 
% hydro
According to the model, run off river hydro generation, which is almost exclusively located in southern Germany, mitigates congestion risk. 
The model seems to strongly overestimate its impact though, probably because of its correlation with hydropower generation in parts of the alpine region.

% SOLAR
Solar generation has a surprisingly low impact and does not mitigate congestions even though it is primarily located in southern Germany. Due to the very strong correlation between generation in the different control areas it is not possible to clearly discern the impact of generation in the different areas. Nevertheless, high solar generation in the 50Hertz control area seems to increase redispatch volumes, as one would expect. Interestingly the model does not show a mitigating effect for solar generation in southern Germany. This might be an effect of the high correlation: While high solar generation in the south decreases congestion due to lower residual loads, this effect might be negated by the simultaneous increased solar generation in northern Germany. If, for instance, the conventional generation that is pushed out of the market, has a similar geographic distribution as the solar generation that replaces it, there is no strong effect on power flows and thus on congestion.

% cross-border flows
Furthermore, imports and exports increase or decrease the likelihood of congestions depending on the location. Especially cross-border flows with Denmark and France have a significant impact on the redispatch volume. Imports in the north from Denmark increase the likelihood of congestions while imports in the south, from France, decrease it and vice versa for exports. 
Imports from Denmark on their own however don't lead to congestion. Only when German wind generation is also significant do they have a negative impact, see Fig.~\ref{fig:wind_flows_vol}. 

Our analysis has further shown that imports from Denmark do in general not fall when wind generation in northern Germany increases.
This is probably due to the limitations of the current market design. Germany and Luxembourg constitute a single bidding zone and the EUPHEMIA algorithm models this bidding zone as a ``copper plate'' with infinite transmission capacity. It therefore regularly calculates dispatches that lead to congestions in Germany, by allowing imports from Denmark also when the German transmission grid is already highly loaded from German wind power.

% Outlook: Redispatch 2.0
As the costs of congestion management increase, the topic is gaining public and political interest. Changes in the regulatory framework are being discussed.
A first step is the establishment of redispatch 2.0 after the Netzausbaubeschleuningungsgesetz (``grid extension law'') has been passed in 2019 and has been in full operation since June 2022 \cite{BundesnetzagenturNetzengpassmanagement} after a three months test phase. A main goal was to include smaller generating units and renewable generation plants in the redispatch scheme. However, our data analysis shows no abrupt changes during the period of study.

% Euphemias role
In the long run, more comprehensive changes in the market design are likely. In particular splitting Germany into a Northern and Southern bidding zone is being discussed. The split would make EUPHEMIA respect the transmission limits between northern and southern Germany, thus reducing congestion from the start. It would also lead to different price levels in the two zones, especially in times of high wind generation~\cite{egerer2016two,fraunholz2021long}.
In the short term, this can strongly affect cross-border electricity trading. 
Lower prices in the Northern zone  would probably decrease imports from Denmark. We have shown in our analysis that imports often remain high even in times of high wind power generation and high congestion in the German grid. In this period, imports will possibly decrease relieving Germany's grid and limiting Denmark's export options.
At the same time, higher prices in the Southern zone would decrease exports to southern Europe.
On longer time scales, the new market situation will lead to new investment incentives, such as for new generation capacity in southern Germany~\cite{fraunholz2021long} or incentives for Power-To-Hydrogen infrastructures~\cite{breder2022spatial}.

The bidding zone split is heavily debated in German politics. The EU’s Agency for the Cooperation of Energy Regulators (ACER) has recently proposed splitting configurations for the German bidding zone~\cite{ACER_2022}. While the northern German states are in favor of the split, six southern German states — for whom the split would presumably lead to higher electricity prices — recently reiterated their opposition~\cite{GemeinsamerBeschlussEnergie2023}.
Notably, the state of Bavaria in Southern Germany hindered the extension of wind power for many years~\cite{langer2016qualitative} and the urgently needed grid expansion is meeting strong opposition in southern Germany~\cite{neukirchGrindingGridContextualizing2020}.

At the same time, other costs related to the energy transition are gaining political interest. Recently, the Federal Minister for Economic Affairs and Climate Action, Robert Habeck, announced an initiative to reform the electricity grid fees to relieve regions with high (renewable) power generation~\cite{zeit2023habeck}.

\section*{Acknowledgments}

This work was performed as part of the Helmholtz School for Data Science in Life, Earth and Energy (HDS-LEE).
We gratefully acknowledge funding from the Helmholtz Association Initiative and Networking Fund through Helmholtz AI.

\bibliographystyle{model1-num-names}
\bibliography{references_manual}

\end{document}